\documentstyle[12pt,aaspp4]{article}

\begin{document}

\vspace*{-1.0in}
\hfill{Accepted for publication in the Astronomical Journal, February 2000}

\title {The Hubble Deep Field South -- STIS Imaging\footnote[1]{Based
on observations made with the NASA/ESA {\em Hubble Space Telescope},
obtained from the Space Telescope Science Institute, which is
operated by the Association of Universities for Research in Astronomy,
Inc., under NASA contract NAS 5-26555.}.}

\author {Jonathan P. Gardner\altaffilmark{2}, 
Stefi A. Baum\altaffilmark{3},
Thomas M. Brown\altaffilmark{2,7},
C. Marcella Carollo\altaffilmark{4,8,9},
Jennifer Christensen\altaffilmark{3},
Ilana Dashevsky\altaffilmark{3},
Mark E. Dickinson\altaffilmark{3},
Brian R. Espey\altaffilmark{3,10},
Henry C. Ferguson\altaffilmark{3},
Andrew S. Fruchter\altaffilmark{3},
Anne M. Gonnella\altaffilmark{3},
Rosa A. Gonzalez-Lopezlira\altaffilmark{3},
Richard N. Hook\altaffilmark{5},
Mary Elizabeth Kaiser\altaffilmark{2,4},
Crystal L. Martin\altaffilmark{3,8},
Kailash C. Sahu\altaffilmark{3},
Sandra Savaglio\altaffilmark{3,10},
T. Ed Smith\altaffilmark{3},
Harry I. Teplitz\altaffilmark{2,7},
Robert E. Williams\altaffilmark{3},
Jennifer Wilson\altaffilmark{3,11}
\\
\altaffiltext{2}{Laboratory for Astronomy and Solar Physics,
Code 681, Goddard Space Flight Center, Greenbelt MD 20771}
\altaffiltext{3}{Space Telescope Science Institute,
3700 San Martin Drive, Baltimore MD 21218}
\altaffiltext{4}{Dept. of Physics and Astronomy, Johns Hopkins University,
Baltimore MD 21218}
\altaffiltext{5}{Space Telescope-European Coordinating Facility, Karl
Schwarzschild Strasse 2, D-85748, Garching bei M\"{u}nchen, Germany}
\altaffiltext{6}{European Southern Observatory, Karl-Schwarzschild-Strasse 2,
D-85748 Garching bei M\"{u}nchen, Germany}
\altaffiltext{7}{NOAO Research Associate}
\altaffiltext{8}{Hubble Fellow}
\altaffiltext{9}{Currently at Columbia University, Department of
Astronomy, Mail Code 5246 Pupin Hall, 550 West 120th Street, New York
NY 10027}
\altaffiltext{10}{On assignment from the Astrophysics Division, Space Science
Department, European Space Agency}
\altaffiltext{11}{Currently at The Observatories of the Carnegie
Institution of Washington, 813 Santa Barbara, Pasadena, CA 91101}
}

\begin{abstract}

We present the imaging observations made with the Space Telescope
Imaging Spectrograph of the Hubble Deep Field -- South. The field
was imaged in 4 bandpasses: a clear CCD bandpass for 156~ksec,
a long-pass filter for 22--25~ksec per pixel typical exposure,
a near-UV bandpass for 23~ksec, and a far-UV bandpass for 52~ksec. 
The clear visible image is the deepest observation ever made in
the UV-optical wavelength region, reaching a $10\sigma$ AB magnitude
of 29.4 for an object of area 0.2 square arcseconds. The field
contains QSO J2233-606, the target of the STIS spectroscopy, and
extends $50\arcsec \times 50\arcsec$ for the visible images, and
$25\arcsec \times 25\arcsec$ for the ultraviolet images. We present
the images, catalog of objects, and galaxy counts obtained in the
field.

\end{abstract}

%\keywords{
%cosmology: observations ---
%galaxies: evolution ---
%galaxies: statistics ---
%surveys
%}
%
\section{Introduction}

The Space Telescope Imaging Spectrograph (STIS) (Kimble et al.\ \markcite{kimble97}1997;
Woodgate et al.\ \markcite{woodgate98}1998; Walborn \& Baum \markcite{walborn98}1998) was used during the Hubble
Deep Field -- South (HDF--S) (Williams et al.\ \markcite{williams99}1999) observations
for ultraviolet spectroscopy (Ferguson et al.\ \markcite{ferguson99}1999) and ultraviolet
and optical imaging. In this paper we present the imaging data.

The Hubble Deep Field -- North (HDF--N) (Williams et al.\ \markcite{williams96}1996) is the
best studied field on the sky, with $>$1~Msec of Hubble Space
Telescope (HST) observing time (including follow-up observations
by Thompson et al.\ \markcite{thompson99}1999 and Dickinson et al.\ \markcite{dickinson99}1999), and countless
observations with ground-based telescopes (e.g., Cohen et al.\ \markcite{cohen96}1996;
Connolly et al.\ \markcite{connolly97}1997). Results obtained to date include a measurement
of the ultraviolet luminosity density of the universe at $z>2$
(Madau et al.\ \markcite{madau96}1996), the morphological distribution of faint galaxies
(Abraham et al.\ \markcite{abraham96}1996), galaxy-galaxy lensing (Hudson et al.\ \markcite{hudson98}1998), and
halo star counts (Elson, Santiago \& Gilmore \markcite{elson96}1996). See Ferguson \markcite{ferguson98}(1998) and
Livio, Fall \& Madau \markcite{livio98}(1998) for reviews and further references. The HDF--S
differs from the HDF--N in several ways. First, the installation
of STIS and NICMOS on HST in 1997 February has enabled parallel
observations with three cameras. In addition to the STIS data, the
HDF--S dataset includes deep WFPC2 imaging (Casertano et al.\ \markcite{casertano99}1999),
deep near-infrared imaging (Fruchter et al.\ \markcite{fruchter99}1999), and wider-area
flanking field observations (Lucas et al.\ \markcite{lucas99}1999). Second, the STIS
observations were centered on QSO J2233-606, at $z \approx 2.24$,
to obtain spectroscopy. Finally, the field was chosen in the southern
HST continuous viewing zone in order to enable follow-up observations
with ground-based telescopes in the southern hemisphere.

In section 2 we describe the observations. In section 3 we describe
the techniques we used to reduce the CCD images. In section 4 we
describe the reduction of the MAMA images. In section 5 we describe
the procedures used to catalog the images. In section 6 we present
some statistics of the data, including galaxy number counts and color
distributions. Our purpose in this paper is to produce a useful
reference for detailed analysis of the STIS images. Thus for the
most part we refrain from model comparisons and speculation on the
significance of the results. We expect the STIS images to be useful
for addressing a wide variety of astronomical topics, including
the sizes of the faintest galaxies, the ultraviolet-optical color
evolution of galaxies, the number of faint stars and white dwarfs
in the galactic halo, and the relation between absorption line
systems seen in the QSO spectrum and galaxies near to the line of
sight. We also expect the observations to be useful for studying
sources very close to the quasar, and perhaps for detecting the
host galaxy of the quasar. However, this may require a re-reduction
of the images, as the quasar is saturated in all of the CCD exposures,
and there are significant problems with scattered light and
reflections.

\section{Description of the observations}

\begin{figure}

\vspace*{-0.2in}

\noindent {\bf NOTE: The resolution of this image has been reduced.
Full resolution images are available at:
http://hires.gsfc.nasa.gov/$\sim$gardner/hdfs/stispaper.}

\vspace{0.5in}

\plotone{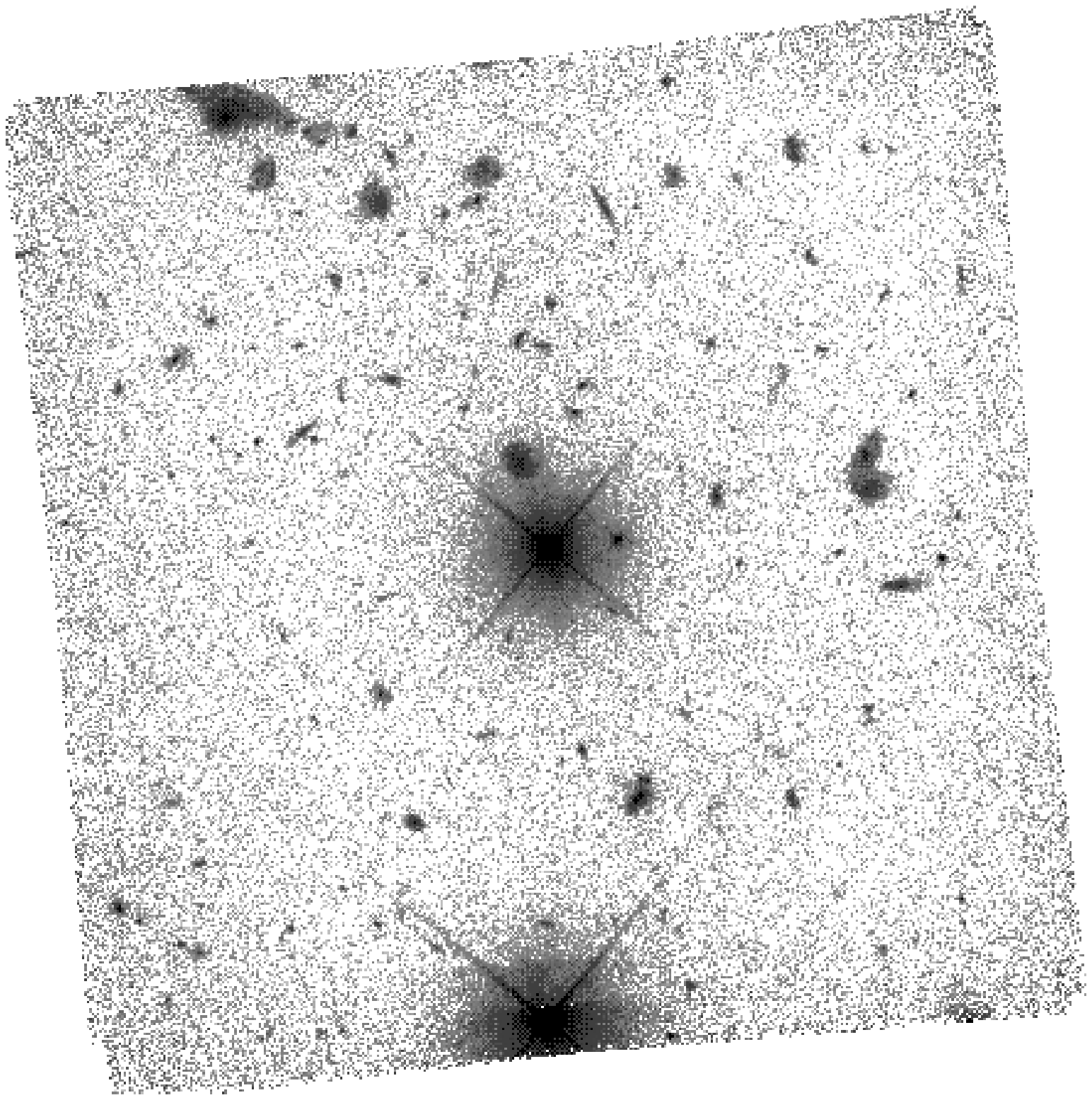}

\caption{The 50CCD image. The image is displayed on a log scale,
and has been clipped between $1 \times 10^{-5}$ and $5 \times
10^{-2}$ counts per second. The field of view of the image is
0.8357 square arcminutes.}

\label{logclr}

\end{figure}

\begin{figure}

\vspace*{-0.2in}

\noindent {\bf NOTE: The resolution of this image has been reduced.
Full resolution images are available at:
http://hires.gsfc.nasa.gov/$\sim$gardner/hdfs/stispaper.}

\vspace{0.5in}

\plotone{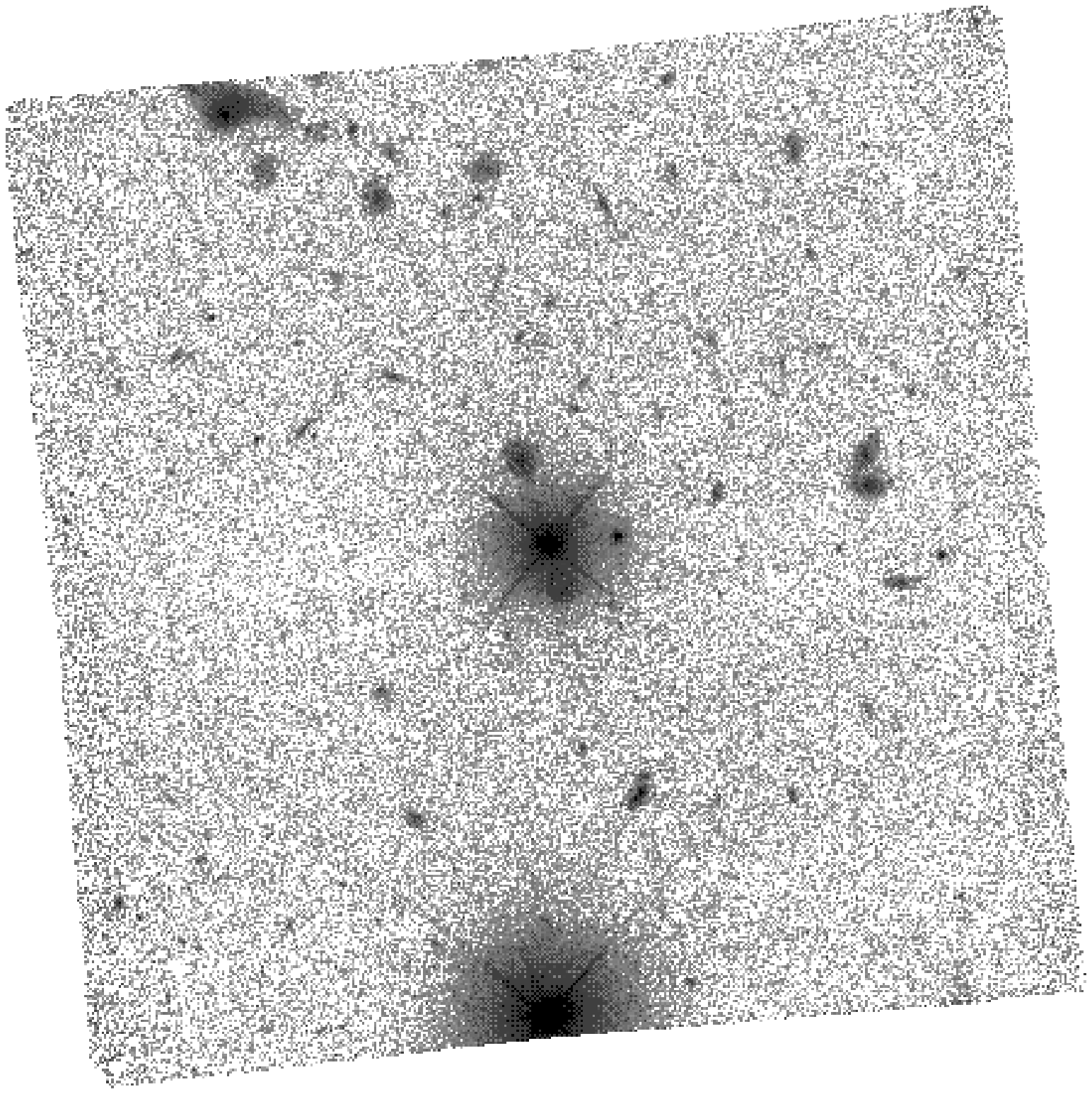}

\caption{The F28X50LP image. The image is displayed on a log scale,
and has been clipped between $1 \times 10^{-5}$ and $5 \times
10^{-2}$ counts per second. The field of view of the image is
0.8326 square arcminutes.}

\label{loglp}

\end{figure}

\begin{figure}

\vspace*{-0.2in}

\noindent {\bf NOTE: The resolution of this image has been reduced.
Full resolution images are available at:
http://hires.gsfc.nasa.gov/$\sim$gardner/hdfs/stispaper.}

\vspace{0.5in}

\plotone{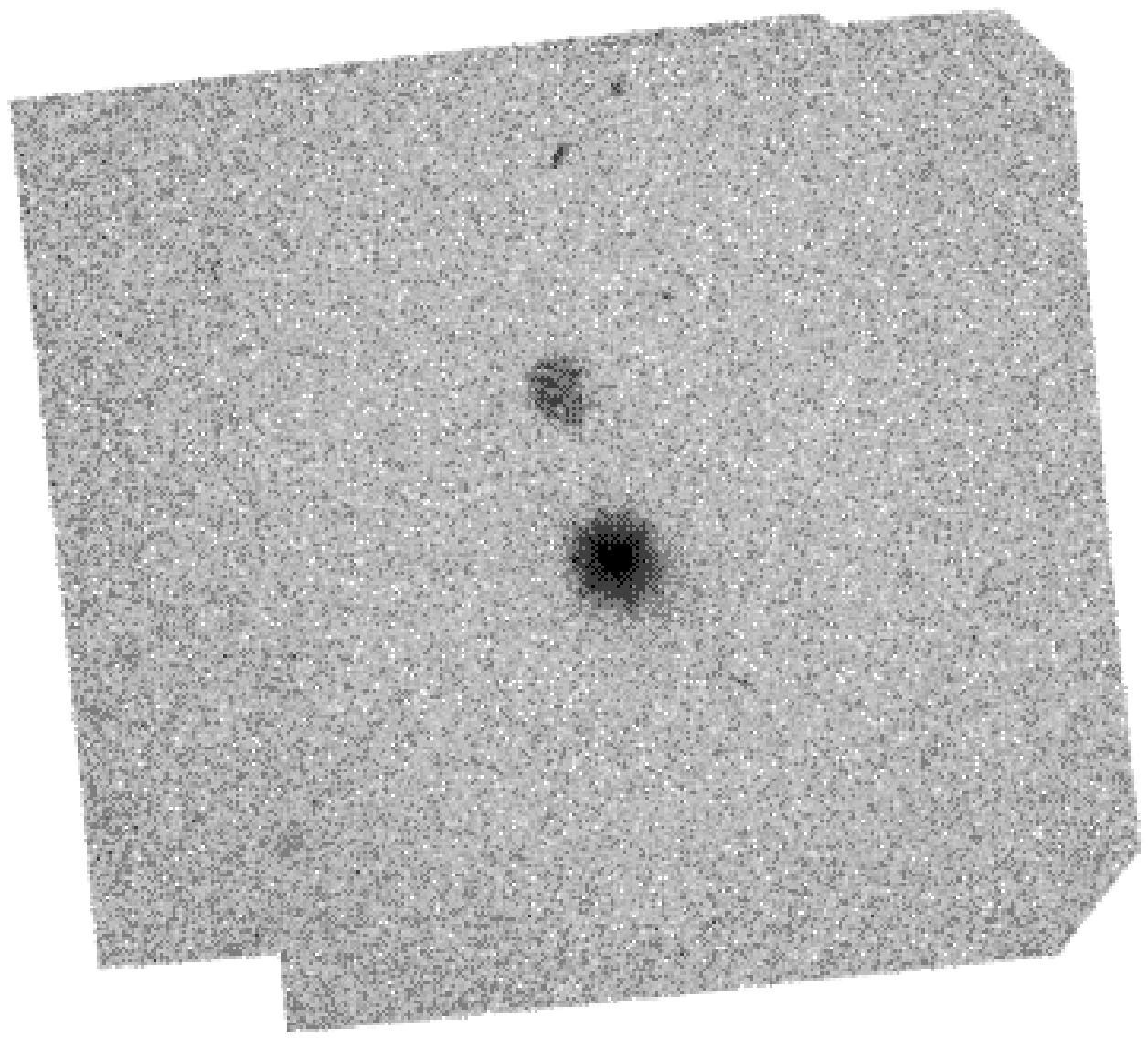}

\caption{The NUVQTZ image. The image is displayed on a log scale,
and has been clipped between $1 \times 10^{-6}$ and $5 \times
10^{-3}$ counts per second, and has been smoothed with a $5 \times
5$ pixel box average. The field of view of the image is
0.2221 square arcminutes.}

\label{lognuv}

\end{figure}

\begin{figure}

\vspace*{-0.2in}

\noindent {\bf NOTE: The resolution of this image has been reduced.
Full resolution images are available at:
http://hires.gsfc.nasa.gov/$\sim$gardner/hdfs/stispaper.}

\vspace{0.5in}

\plotone{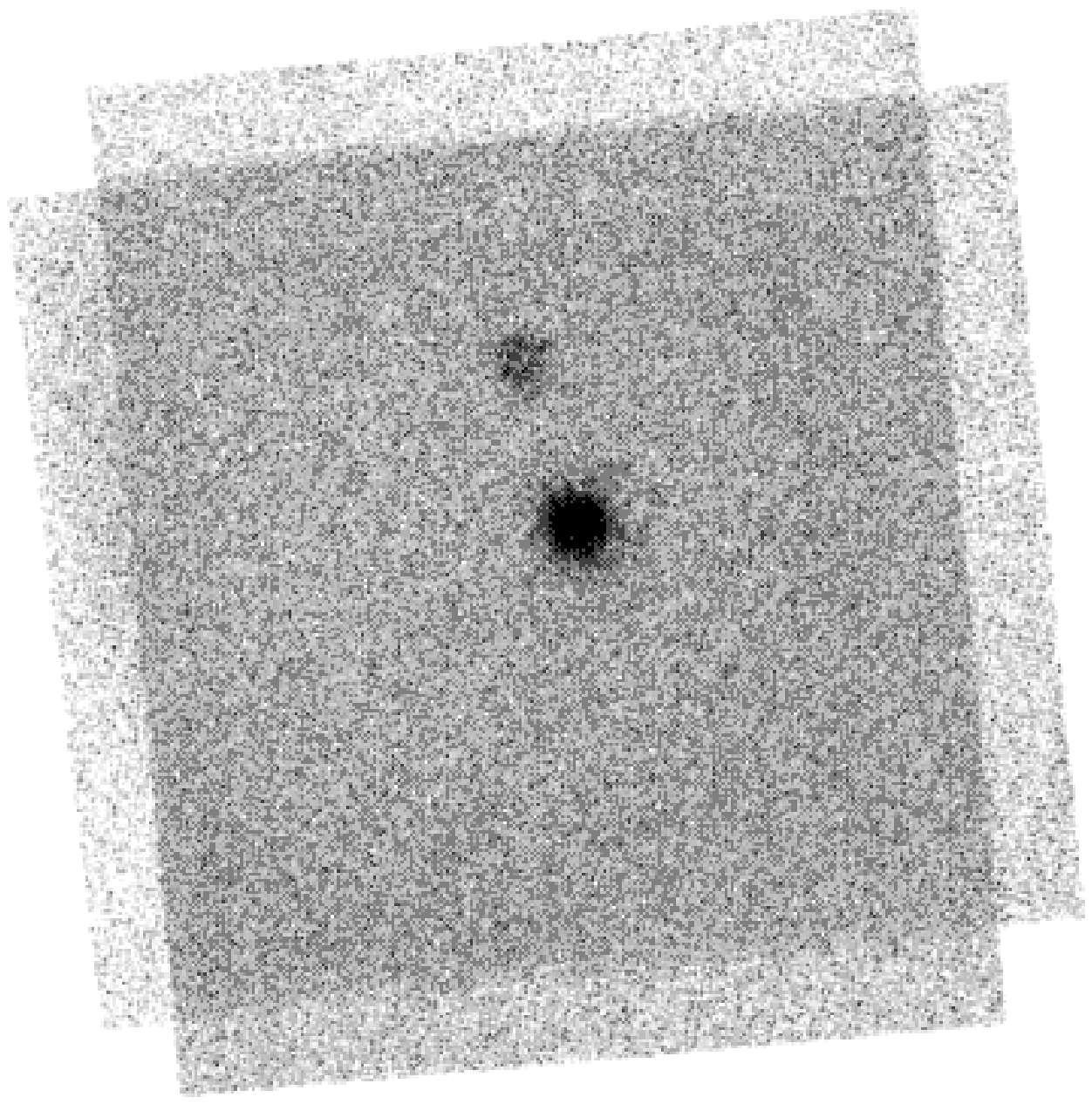}

\caption{The FUVQTZ image. The image is displayed on a log scale,
and has been clipped between $1 \times 10^{-8}$ and $5 \times
10^{-5}$ counts per second, and has been smoothed with a $5 \times
5$ pixel box average. The field of view of the image is
0.2438 square arcminutes.}

\label{logfuv}

\end{figure}

\begin{figure}

\plotone{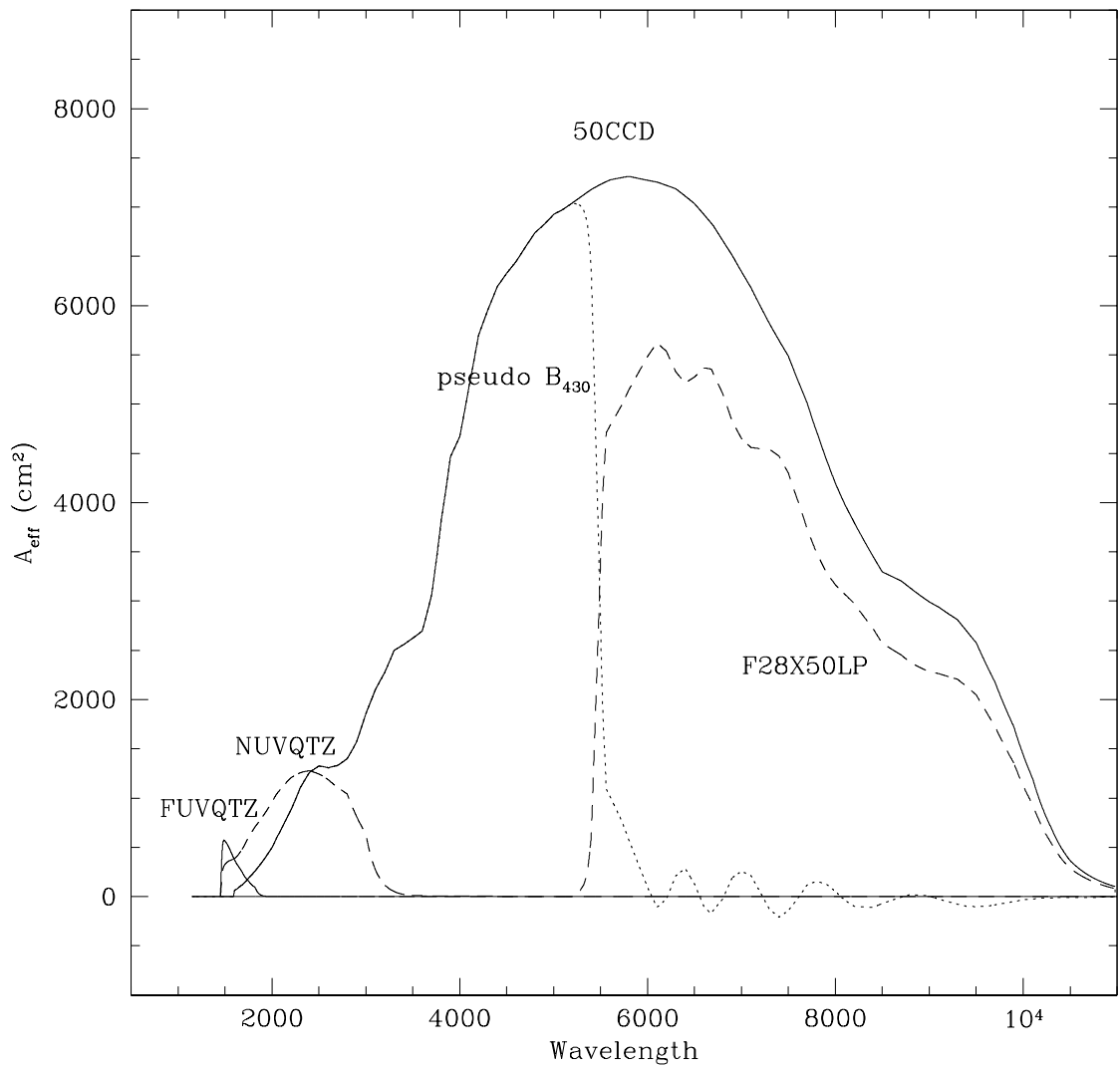}

\caption{Effective areas of the 4 imaging modes. The 50CCD mode is
filterless imaging with a CCD, and this curve represents the response
of the detector. The other three modes are a combination of the
throughput of the filter with the response functions of the CCD
and the two MAMA detectors. Also plotted is a pseudo-$B_{430}$
bandpass, constructed from the fluxes by 50CCD - 1.31(F28X50LP).}

\label{filttrans}

\end{figure}

The images presented here were taken in 4 different modes, 50CCD
(Figure~\ref{logclr}), F28X50LP (Figure~\ref{loglp}), NUVQTZ
(Figure~\ref{lognuv}), and FUVQTZ (Figure~\ref{logfuv}). The 50CCD
and F28X50LP modes used the Charge Coupled Device (CCD) detector.
The 50CCD is a clear, filterless mode, while the F28X50LP mode uses
a long-pass filter beginning at about 5500{\AA}. The FUVQTZ and
NUVQTZ used the Multi-Anode Microchannel Array (MAMA) detectors as
imagers with the quartz filter. The quartz filter was selected to
reduce the sky noise due to airglow to levels below the dark noise. The
effective areas of the 4 modes are plotted in Figure~\ref{filttrans},
along with a pseudo-$B_{430}$ bandpass constructed from the 50CCD
and F28X50LP fluxes. The MAMA field of view is a square, $25\arcsec$
on a side, and was dithered so that the observations include data
on a field approximately $30\arcsec$ square. The 50CCD mode is
filterless imaging with a CCD. The field of view is a square
$50\arcsec$ on a side, and the dithering extends to a square
$60\arcsec$ on a side. The F28X50LP is a long-pass filter that
vignettes the field of view of the CCD to a rectangle $28\times
50\arcsec$. The observations were dithered to image the entire
field of view of the 50CCD observations, although the exposure time
per point on the sky is thus approximately half the total exposure
time spent in this mode. The original pixel scale is
$0.0244\arcsec$~pix$^{-1}$ for the MAMA images, and
$0.05071\arcsec$~pix$^{-1}$ for the CCD images. The final combined
images have a scale of $0.025\arcsec$~pix$^{-1}$ in all cases.
Table~\ref{obstab} describes the observations. The filterless 50CCD
observations correspond roughly to V+I, and reach a depth of 29.4
AB magnitudes at $10\sigma$ in a 0.2 square arcsecond aperture (320
drizzled pixels). This is the deepest exposure ever made in the
UV-optical wavelength region.

\subsection{Selection of the Field}

Selection of the field is described by Williams et al.\ \markcite{williams99}(1999). The QSO
is at RA~=~$\rm 22^h33^m37.5883^s$, Dec~=~$-60^{\circ} 33\arcmin
29.128\arcsec$ (J2000). The errors on this position are estimated
to be less than 40 milli-arcseconds (Zacharias et al.\ \markcite{zacharias98}1998). The
position of the QSO on the 50CCD and F28X50LP images is x=1206.61,
y=1206.32, and on the MAMA images is x=806.61, y=806.32.

\subsection{Test Data}

Test observations of the field were made in 1997 October. 
These data are not used in the present analysis. While the test
exposures do not add significantly to the exposure time, they would
provide a one-year baseline for proper motion studies of the brighter
objects.

\subsection{Observing Plan}

The STIS observations were scheduled so that the CCD was used in the
orbits that were impacted by the South Atlantic Anomaly, and the
MAMAs were used in the clear orbits. The observations were made in the
continuous viewing zone, and therefore were all made close to the
limb of the Earth. The G430M spectroscopy, all of which was read-noise
limited, was done during the day or bright part of the orbit, while
the CCD imaging was all done during the night or dark part of the
orbit. The MAMA imaging, done with the quartz filter, is insensitive
to scattered Earth light, and was therefore done during bright time. A
more detailed discussion of the scheduling issues is given by
Williams et al.\ \markcite{williams99}(1999). The sky levels in the 50CCD images were
approximately twice the square of the read noise, so these data are
marginally sky noise limited. The MAMA images are limited by the dark
noise.

\subsection{Dithering and Rotation}

The images were dithered in right ascension (RA) and declination
(Dec) in order to sample the sky at the sub-pixel level. In addition,
variations in rotation of about $\pm 1$ degree were used to provide
additional dithering for the WFPC2 and NICMOS fields during the
STIS spectroscopic observations. The STIS imaging observations were
interspersed with the STIS spectroscopic observations; therefore,
all of the images were dithered in rotation as well as RA and Dec.

\subsection{CR-SPLIT and pointing strategy}

The CCD exposures were split into 2 or 3 {\sc cr-split}s that each
have the same RA, Dec, and rotation. This facilitates cosmic ray
removal, although as discussed below, this was only used in the
first iteration of the data reduction. The final 50CCD image is
the combination of 193 exposures making up 67 {\sc cr-split}
pointings. After standard pipeline processing, (including bias and
dark subtraction, and flatfielding), each exposure is given a {\sc
flt} file extension, and the cosmic-ray rejected combinations of
each {\sc cr-split} is given a {\sc crj} file extension. The final
F28X50LP image is the combination of 66 exposures making up 23 {\sc
cr-split} pointings. The F28X50LP image included 12 pointings at
the northern part of the field, one pointing at the middle of the
field, and 10 pointings at the southern half of the field.

\subsection{PSF observations}

In order to allow for PSF subtraction of the QSO present in the
center of the STIS 50CCD image, two SAO stars of about 10~mag were
observed in the filterless 50CCD mode before and after the main
HDF-S campaign. The stars are SAO 255267, a G2 star, and SAO 255271,
an F8 star, respectively. These targets have spectral energy
distributions in the STIS CCD sensitivity range similar to that of
the QSO. For each star, 32 different {\sc cr-split} exposures were taken.
The following strategy was used: (i) four different exposure times
between 0.1 s and 5 s for each {\sc cr-split} frame, to ensure high
signal-to-noise in the wings while not saturating the center; (ii)
a four-position dither pattern with quarter-pixel sampling and
{\sc cr-split} at each pointing with each exposure time; (iii) use of
gain=4, to insure no saturation in the A-to-D conversion. During
the observations for SAO255267, a failure in the guide star
acquisition procedure caused the loss of its long-exposure (5~s)
images. Gain=4 has a well-documented large scale pattern noise that
must be removed, e.g., by Fourier filtering, before a reliable PSF
can be produced. These data are not discussed further in this paper,
but are available from the HST archive for further analysis.

\section{Reduction of the CCD Images}

\subsection{Bias, Darks, Flats and Masks}

Standard processing of CCD images involves bias and dark subtraction,
flatfielding, and masking of detector defects. The bias calibration
file used for the HDF-S was constructed from 285 individual exposures,
combined together with cosmic-ray and hot-pixel trail rejection.

The dark file was constructed from a ``superdark'' frame and a
``delta'' dark frame. The superdark is the cosmic-ray rejected
combination of over 100 individual 1200~s dark exposures taken
over the several months preceding the HDF-S campaign. The delta
dark adds into this high S/N dark frame the pixels that are more
than $5\sigma$ from the mean in the superdark-subtracted combination
of 14 dark exposures taken during the HDF-S campaign. Calibration
of the images with this dark frame removes most of the hot pixels
but still leaves several hundred in each image.

An image mask was constructed to remove the remaining hot pixels
and detector features. The individual cosmic-ray rejected HDF-S
50CCD exposures were averaged together without registration. The
remaining hot pixels were identified with the IRAF\footnote[12]{IRAF
is distributed by the National Optical Astronomy Observatories,
which are operated by the Association of Universities for Research
in Astronomy, Inc., under cooperative agreement with the National
Science Foundation.} {\sc cosmicrays} task. These pixels were
included in a mask that was used to reject pixels during the
{\sc drizzle} phase. Pixels that were more than $5\sigma$ below the mean
sky background were also masked, as were the 30 worst hot pixel
trails, and the unilluminated portions of the detector around the
edges. Hot pixel trails run along columns and are caused by high
dark current in a single pixel along the column.

Flatfielding was carried out by the IRAF/STSDAS {\sc calstis}
pipeline using two reference files. The first, the {\sc pflat}
corrects for small-scale pixel-to-pixel sensitivity variations,
but is smooth on large scales. This file was created from ground-test
data but comparisons to a preliminary version of the on-orbit flat
revealed only a few places where the difference was more than 1\%.
The CCD also shows a 5-10\% decrease in sensitivity near the edges
due to vignetting. This illumination pattern was corrected by a
low-order fit to a sky flat constructed from the flanking field
observations.

\subsection{Shifts and rotations}

After pipeline processing, the CCD images were reduced using the
IRAF/STSDAS package {\sc dither}, and test versions called {\sc
xdither}, and {\sc xditherii}. These packages include the {\sc
drizzle} software (Fruchter \& Hook \markcite{fruchterhook98}1998; Fruchter et al.\ \markcite{fruchteretal98}1998;
Fruchter \markcite{fruchter98}1998). We used {\sc drizzle} version 1.2, dated 1998
February. The test versions differ from the previously released
version primarily in their ability to remove cosmic rays from each
individual exposure, and include tasks that have not yet been
released.

The {\sc xditherii} package uses an iterative process to reject cosmic
rays and determine the x and y sub-pixel shifts, which we summarize
here. The standard pipeline rejects cosmic rays using each {\sc
cr-split} of 2 or 3 images. The resulting {\sc crj} files are used
as the first iteration, we determine the x and y shifts, and the
files are median combined. The resulting preliminary combination
is then shifted back into the frame of each of the original exposures
({\sc flt} files), and a new cosmic ray mask is made. By comparing
each exposure to a high signal-to-noise combination of all of the
data, we are less likely to leave cosmic ray residuals. The x and
y shifts are determined at each iteration as well.

The rotations used in combining the data were determined from the
{\sc roll\_avg} parameter in the jitter files, using the program
{\sc bearing}. We did not seek to improve on these rotations via
cross-correlation or any other method. We did use cross-correlation
to determine the x and y shifts.

Determination of the sub-pixel x and y shifts was done with an
iterative procedure. The first iteration was obtained by determining
the centroid of the bright point source just west of the QSO, using
the pipeline cosmic-ray rejected {\sc crj} files. We could not use
cross-correlation in this first iteration, because the very bright
star on the southern edge of the field was present on images taken
at some, but not all, dither positions, which corrupted the
cross-correlation. The source we used for centroiding was clearly
visible on all of the 50CCD and F28X50LP frames.

Using these shifts (which were accurate to better than 1 pixel),
we created a preliminary combined image. After pipeline processing
and cosmic ray rejection, the {\sc drizzle} program was used to
shift and rotate each {sc crj} file onto individual outputs, without
combining them. We then used the task {\sc imcombine} to create a
median combination of the files. This preliminary image was then
shifted and rotated back into the frame of each individual exposure
using the {\sc xdither} task, {\sc blot}, ready for the next iteration
of the cosmic-ray rejection procedure. 

\subsection{Cosmic ray rejection}

In this iteration, we discarded the {\sc crj} files, and went back
to the {\sc flt} files, in which each exposure had undergone bias
and dark subtraction and flatfielding, but not cosmic-ray rejection.
Each exposure was compared to the blotted image, and a cosmic-ray
mask for that exposure was created from all of the pixels that
differed (positively or negatively) by more than a given threshold
from the blotted image. In the version 1.0 released 50CCD image,
this threshold was set to be $5\sigma$. However, we believe that
a small error in the sky level determination, introduced by the
amplifier ringing correction discussed below, meant that our 
rejection was approximately at the $3\sigma$ level.
The cosmic ray masks were multiplied by the hot pixel masks discussed
above, and resulted in about 8\% of the pixels being masked as
either cosmic rays or hot pixels. This is, perhaps, overly
conservative. A less conservative cut (after correcting the error
in the sky value) would result in slightly higher exposure time
per pixel, and thus an improvement of 1-2\% in the signal to noise
ratio. The cosmic ray mask was combined with the hot pixel and
cosmetic defect mask.

This problem with the sky value was corrected in the F28X50LP image,
and a $3\sigma$ level was used in the cosmic ray rejection.

\subsection{Amplifier ringing correction}

Horizontal features due to amplifier ringing, varying in pattern
from image to image, were present in most of the STIS CCD frames.
When a pixel saw a highly saturated signal, the bias level was
depressed in the readout for the next few rows. The very high
signals causing this ringing came from hot pixels and from the
saturated QSO. The signal-to-noise ratio in the overscan region of
the detector was not sufficient to remove these features well. We
removed them with a procedure that subtracted on a row-by-row basis,
from each individual image, the weighted average of the background
as derived from the innermost 800 columns after masking and rejecting
``contaminated'' pixels. The masks included all visible sources,
hot pixels, and cosmic-ray hits. The source mask was determined
from the initial registered median-combined image, shifted back to
the reference frame of each of the individual images. For the
unmasked pixels in each row, the 50 highest and lowest were rejected
and the mean of the remaining pixels was subtracted from the each
pixel in that row.

Heavily smoothing the images reveals very slight horizontal residuals
that were not removed by the present choice of parameters in this
process.

\subsection{Drizzling it all together}

\begin{figure}

\plotone{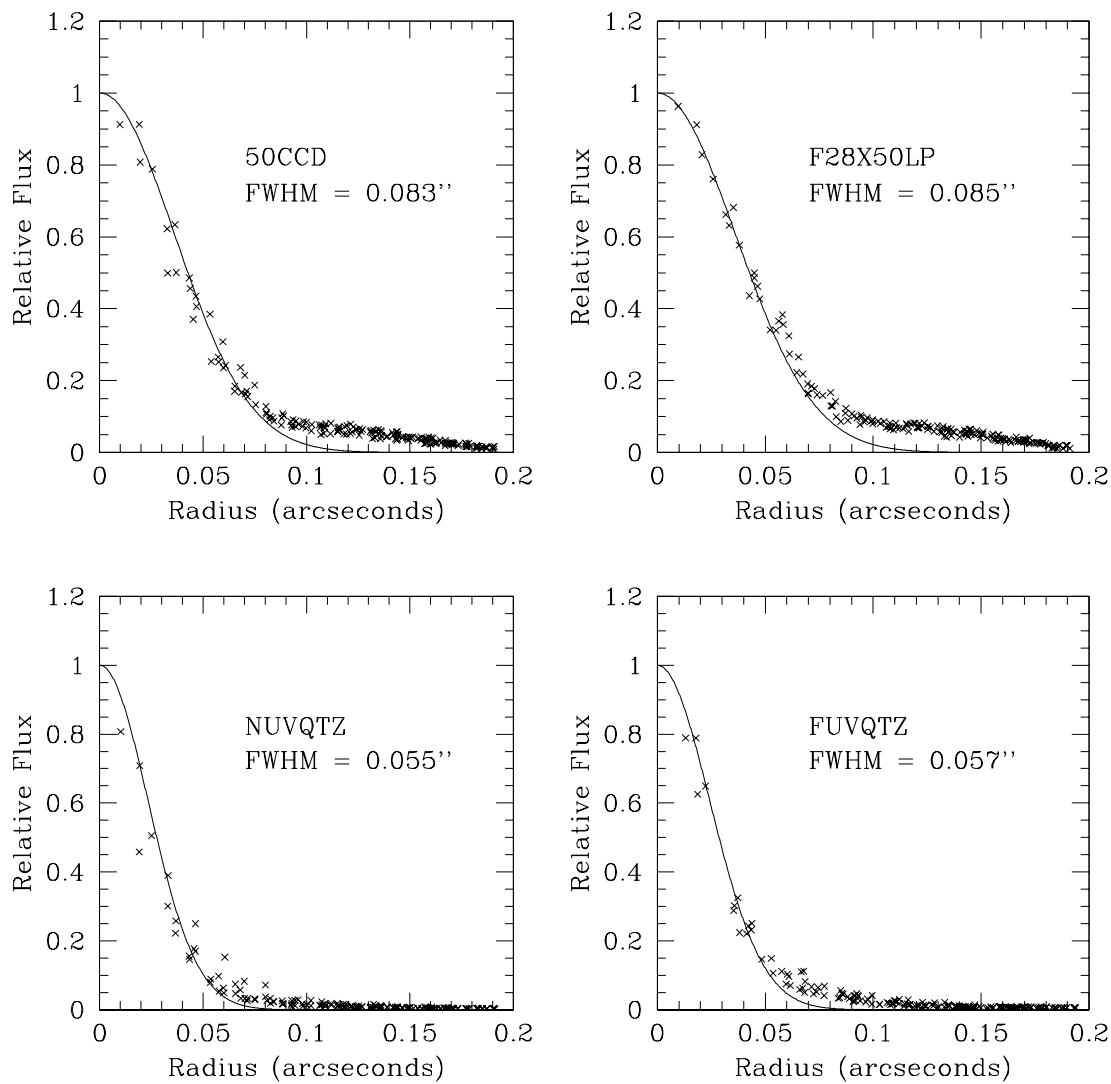}

\caption{Point spread functions of the final images. The points
plotted are each pixel value as a function of distance from the
centroid of the point source. The lines are a Gaussian with the same
full width half maximum as the PSF. The objects plotted are the red
point source just to the west of the quasar in the optical images, and
the quasar itself in the ultraviolet images.}

\label{psf}

\end{figure}

The final image combination was done by drizzling the amplifier-ringing
corrected pipeline products together onto a single output image.
The exposures were weighted by the square of the exposure time,
divided by the variance, which is (sky+rn$^2$+dark). The rotations
were corrected so that North is in the +y direction, and the scale
used was 0.492999 original CCD pixels per output pixel so that the
final pixel scale is exactly 0.025 arcsec/pixel. For the 50CCD data
we used a {\sc pixfrac}=0.1, which is approximately equivalent to
interleaving, where each input pixel falls on a single output pixel.
For the F28X50LP data we used {\sc pixfrac}=0.6, as a smaller {\sc
pixfrac} left visible holes in the final image. See Fruchter \& Hook \markcite{fruchterhook98}(1998)
for a discussion of the meaning of the {\sc drizzle} parameters. The
point spread functions of bright, non-saturated point sources are
shown in Figure~\ref{psf}. The sources selected are the point source
just to the west of the quasar in the 50CCD and F28X50LP images,
and the QSO in the MAMA images.

The final image is given in counts per second, which can be converted
to magnitudes on the {\sc stmag} system using the photometric
zeropoints given by the {\sc photflam} parameter supplied in the
image headers. We used the pipeline photometric zeropoints for the
50CCD  and MAMA images, but revised the F28X50LP zeropoint by 0.1
magnitude based on a comparison of STIS photometry of the HST
calibration field in $\omega$ Centauri with the ground-based
photometry of Walker \markcite{walker94}(1994). The zeropoints in the AB magnitude
system which we used are 26.386, 25.291, 23.887, and 21.539, for
the 50CCD, F28X50LP, NUVQTZ and FUVQTZ respectively. We also supply
the weight image, which is the sum of the weights falling on each
pixel. For the F28X50LP image, we supply an exposure-time image,
which is the total exposure time contributing to each pixel. We
have multiplied this image by the area of the output pixels. The
world coordinate system in the headers was corrected so that North
is exactly in the +y direction, and the pixel scale is exactly
0.025 arcsec/pixel.

\subsection{Window reflection}

A window in the STIS CCD reflects slightly out-of-focus light from
bright sources to the +x, $-$y direction (SW on the HDF-S images).
The QSO is saturated in every 50CCD and F28X50LP exposure. The
window reflection of the QSO is clearly visible in the F28X50LP
image, but has been partially removed from the 50CCD image by the
cosmic-ray rejection procedure. We wish to emphasize that it has
only been partially removed, and there are remaining residuals.
These residuals should not be mistaken for galaxies near the QSO,
nor should they be mistaken for the host galaxy of the QSO. There
is additional reflected light from the QSO (and from the bright
star at the southern edge) evident in the images. We believe that
the version 1.0 released images are not appropriate for searching
for objects very close to or underlying the QSO, and that such a
search would require re-processing the raw data with particular
attention paid to the window reflection, other reflected light,
and to the PSF of the QSO. The diffraction spikes of the QSO are
smeared in the final images by the rotation of the individual
exposures.

\section{Reduction of the MAMA Images}

The near-UV and far-UV images are respectively the weighted averages
of 12 and 25 registered frames, with total exposure times of 22616~s
and 52124~s. The MAMAs do not suffer from read noise or cosmic
rays, and the quasar is not saturated in any of the UV data. However,
the MAMAs do have calibration issues that must be addressed.

\subsection{Flats, Dark Counts, and Geometric Correction}

Prior to combination, all frames were processed with CALSTIS,
including updated high-resolution pixel-to-pixel flat field files
for both UV detectors. Geometric correction and rescaling were
applied in the final combinations via the {\sc drizzle} program. The
quartz filter changes the far-UV plate scale relative to that in
the far-UV clear mode, and so the relative scale between MAMA
imaging modes was determined from calibration images of the globular
cluster NGC~6681.

Dark subtraction for the near-UV image was done by subtracting a
scaled and flat-fielded dark image from each near-UV frame. The
scale for the dark image was determined by inspection of the
right-hand corners of the near-UV image, because these portions of
the detector are occulted by the aperture mask and thus only register
dark counts. For the far-UV images, {\sc calstis} removes a nearly flat
dark frame, but the upper left-hand quadrant of STIS far-UV frames
contains a residual glow in the dark current after nominal calibration.
This glow varies from frame to frame and also appears to change
shape slightly with time. To remove the residual dark current, the
16 far-UV frames with the highest count rates in the glow region
were co-added without object registration but with individual object
masks for the only two obvious objects in the far-UV frames (the
quasar and bright spiral NNE of the quasar). We then fit the result
with a cubic spline to produce a glow profile. This profile was
then scaled to the residual glow in each processed frame and
subtracted prior to the final drizzle. Even during observations
with a strong dark glow, where the dark count rate is an order of
magnitude higher than normal, it is still very low, reaching rates
no higher than $6\times 10^{-5}$cts~sec$^{-1}$~pix$^{-1}$. The glow
thus appears as a higher concentration of ones in a sea of zeros,
and the subtraction of a smooth glow profile from such quantized
data over-subtracts from the zeros and under-subtracts from the
ones. These effects are visible in the corrected data, even when
smoothed out considerably in the final drizzled far-UV image. A
low-resolution flat-field correction was applied to the far-UV
frames after subtraction of the residual dark glow. The near-UV
frames require no low-resolution flat field correction.

\subsection{Shifts and Rotations}

Currently, geometrically corrected NUVQTZ and FUVQTZ frames do not
have the same plate scale.  Although geometric correction, rotation,
and rescaling is applied during the final summation of individual
calibrated frames, we first produced a set of calibrated frames that
included these corrections, in order to accurately determine the
relative shifts between them; this information was then used in
conjunction with these corrections in the final drizzle.  All near-UV
and far-UV frames were geometrically corrected, rescaled to
$0.025\arcsec~pix^{-1}$, and rotated to align North with the +y image
axis. The roll angle specified in the jitter files was used to
determine the relative roll between frames, and the mean difference
between the planned roll and the jitter roll determined the absolute
rotation. It is difficult to determine accurate roll angles from the
images themselves, because of the scarcity of objects in the MAMA
images. All near-UV and far-UV frames were then cross-correlated
against one of the far-UV frames to provide shifts in the output
coordinate system. Note that centroiding on the quasar in all far-UV
and near-UV frames yields the same shifts as cross-correlation, within
0.1 pixel.

\subsection{Drizzling}

The calibrated frames were drizzled to a $1600 \times 1600$ pixel
image, including the above corrections, rescaling, rotations, and
shifts. We updated the world coordinate system in the image headers
to exactly reflect the plate scale, alignment, and the astrometry
of the quasar.

For both the far-UV and near-UV frames, individual pixels in each
frame were weighted by the ratio of the exposure time squared to
the dark count variance; this weights the exposures by (S/N)$^2$
for sources that are fainter than the background. Although the
variations in the far-UV dark profile are smooth, the near-UV dark
profile is an actual sum of dark frames, and so we smoothed the
near-UV dark profile to determine the weights. With this weighting
algorithm, pixels in the upper left-hand quadrant of a given far-UV
image contribute less when the dark glow is high, and contribute
more when it is low. The statistical errors (cts~s$^{-1}$) in the
final drizzled image, for objects below the background (e.g.,
objects other than the quasar), are given by the square root of
the final drizzled weights file.

The drizzle ``dropsize'' ({\sc pixfrac}) was 0.6, thus improving the
resolution over a {\sc pixfrac} of 1.0 (which would be equivalent to
simple shift-and-add). The $1600 \times 1600$ pixel format contains
all dither positions, and pixels outside of the dither pattern
are at a count rate of zero. The pixel mask for each near-UV input
frame included the occulted corners of the detector, a small number
of hot pixels, and pixels with relatively low response (those with
values $\le$ 0.75 in the high-resolution flat field). The pixel
mask for each far-UV frame included hot pixels and all pixels
flagged in the data quality file for that frame. When every input
pixel drizzled onto a given output pixel was masked, that
pixel was set to zero.

\subsection{Window Reflection}

As with the CCD, a window reflection of the QSO
appears in the near-UV image. This reflection appears $\approx 0.2\arcsec$
east of the QSO itself, and should not be considered an astronomical object.
 
\section{Cataloging}

\subsection{Cataloging the Optical Images}

The catalog was created using the {\sc SExtractor} package
(Bertin \& Arnouts \markcite{bertin96}1996), revision of 1998 November 19, with some minor
modifications that were done for this application. We used two
separate runs of {\sc SExtractor}, and manually merged the resulting
output catalogs. The first run used a set of parameters selected
to optimize the detection of faint sources while not splitting what
appeared to the eye to be substructure in a single object. We varied
the parameters {\sc detect\_thresh}, {\sc deblend\_mincont}, {\sc
back\_size}, and {\sc back\_filtersize}. We decided to use a
detection threshold corresponding to an isophote of $0.65\sigma$.
Sources were required a minimum area of 16 connected pixels above
this threshold. Deblending was done when the flux in the fainter
object was a minimum of 0.03 times the flux in the brighter object.
The background map was constructed on a grid of 60 pixels, and
subsequently filtered with a $3\times3$ median filter. Prior to
cataloging, the image was convolved with a Gaussian kernel with
full width half maximum of 3.4 pixels. As discussed in
Fruchter \& Hook \markcite{fruchterhook98}(1998), the effects of drizzling on the photometry
is no more than 2\%, and in our well-sampled 50CCD field, the
effects should be much less than this. This effect is smaller than
other uncertainties in the photometry of extended objects.

The second run of {\sc SExtractor} was optimized to detect objects
that lay near the QSO and the bright star at the southern edge of
the image. These objects tend to be blended in with the point source
at the lower detection threshold. Although our catalog might include
galaxies that are associated with absorption lines in the quasar
spectrum, we did not attempt to subtract the quasar light from the
image, and so the catalog does not include objects within $3\arcsec$
of the quasar. The parameters used for the second run were the same
as for the first run, with the exception of the {\sc detect\_thresh}
parameter, which was set to $3.25\sigma$. This parameter not only
sets the minimum flux level for detection, but also is the isophote
used to determine the extent of the object. Several objects fall
between the $0.65\sigma$ isophote and the $3.25\sigma$ isophote of
the quasar. These are not deblended on the first {\sc SExtractor}
run, because their fluxes are below 0.03 of the quasar flux, but
are detected (without the need for deblending) on the second run.
Objects near the quasar detected in the second run were added to
the catalog generated by the first run, and flagged accordingly.
Objects from the second run that were not confused with the quasar
or the bright star were not included. The isophotal photometry of
objects from the second run will not be consistent with the photometry
of objects from the first run, because a different isophote was
used. Eight objects were added to the catalog in this way.

In addition, 26 objects from the first {\sc SExtractor} run 
were clearly spurious due to the diffraction spikes of the QSO and 
the bright star. These were manually deleted from the catalog. 

Photometry of the F28X50LP image was done with {\sc SExtractor}
run in two-image mode, in which the objects were detected and
identified on the 50CCD image, but the photometry was done in the
other band. Isophotes and elliptical apertures are thus determined
by the extent of the objects on the 50CCD images. Objects detected
in the F28X50LP image but not on the 50CCD image are impossible,
since it has a lower throughput and shorter exposure time.

\subsection{Cataloging the Ultraviolet Images}

Fluxes in the UV were calculated outside of {\sc SExtractor} because
it had some problems handling quantized low-signal data. To determine
the gross flux, we summed the countrate within the area for each
object appearing in the {\sc SExtractor} 50CCD segmentation map.
We then created an object mask by ``growing'' each object in the
segmentation map, using the IDL routine {\sc dilate}, until it
subtended an area three times its original size. The resulting mask
excludes faint emission outside of the {\sc SExtractor} isophotes
for all known objects in the field. The sky was calculated from
those exposed pixels within a $151 \times 151$ pixel box centered
on each object, excluding pixels from the mask. The mean countrate
per pixel in this sky region was used to determine the background
for each object (the median is not a useful quantity when dealing
with very low quantized signals), and thus the net flux. Statistical
errors per pixel for objects at or below the background are determined
from the {\sc drizzle} weight image raised to the $-1/2$ power.
The statistical errors for the gross flux and sky flux were calculated
using this pixel map of statistical errors, and thus underestimate
the errors for bright objects such as the quasar.

Some objects that are fully-exposed in the CCD image do not fall
entirely within the exposed area of the MAMA images; for these
objects, we calculated the UV flux in the exposed area only, without
correcting for the incomplete exposure, and flagged such objects
accordingly. Objects were also flagged if the sky-box described
above did not contain at least 100 pixels (e.g., the quasar). For
these objects, we calculated a global sky value from a larger $685
\times 670$ pixel box, roughly centered in each MAMA image, that
only includes areas fully exposed in the dither pattern, and excludes
pixels in the object mask. When the net flux incorporates this
global sky value, they have been flagged accordingly. We do not
expect or see any evidence for objects in the ultraviolet images
that do not appear on the 50CCD image.

\subsection{The Catalog}

The catalog is presented in Table~\ref{cattab}, which contains a
subset of the photometry. The full catalogs are available on the World
Wide Web. For each object we report the following parameters:

{\bf ID:} The {\sc SExtractor} identification number. The objects
in the list have been sorted by right ascension (first) and
declination (second), and thus are no longer in catalog order. In
addition, the numbers are no longer continuous, as some of the object
identifications from the first {\sc SExtractor} run have been
removed. Objects from the second {\sc SExtractor} run have had
10000 added to their identification numbers. These identification
numbers provide a cross-reference to the segmentation maps.

{\bf HDFS\_J22r$-$60d:} The minutes and seconds of right ascension
and declination, from which can be constructed the catalog name of
each object. To these must be added 22 hours (RA) and $-60$ degrees
(Dec). The first object in the catalog is HDFS\_J223333.69$-$603346.0,
at RA 22$^h$ 33$^m$ 33.69$^s$, Dec 60$\deg$ 33$\arcmin$
46.0$\arcsec$, epoch J2000.

{\bf x, y:} The x and y pixel positions of the object on the 50CCD and
F28X50LP images. To get the x and y pixel positions on the MAMA
images, subtract 400 from each.

{\bf $m_i$, $m_a$:} The isophotal ($m_i$) and ``mag\_auto'' ($m_a$)
50CCD magnitudes. The magnitudes are given in the AB system
(Oke \markcite{oke71}1971), where $m = -2.5 log f_{\nu} - 48.60$. The isophotal
magnitude is determined from the sum of the counts within the
detection isophote, set to be 0.65$\sigma$. The ``mag\_auto'' is
an elliptical Kron \markcite{kron80}(1980) magnitude, determined from the sum of
the counts in an elliptical aperture. The semi-major axis of 
this aperture is defined by 2.5 times the first
moments of the flux distribution within an ellipse roughly twice the
isophotal radius. However if the aperture defined this way 
would have a semi-major axis smaller than than 3.5 pixels, a
3.5 pixel value is used.

{\bf clr-lp:} Isophotal color, 50CCD$-$F28X50LP, in the AB magnitude
system, as determined in the 50CCD isophote. {\sc SExtractor} was
run in two-image mode to determine the photometry in the F28X50LP
image, using the 50CCD image as the detection image. When the
measured F28X50LP flux is less than $2\sigma$, we determine an
upper limit to the color using the flux plus $2\sigma$ when the
measured flux is positive, and $2\sigma$ when the measured flux is
negative. We did not clip the 50CCD photometry.

{\bf nuv-clr, fuv-clr:} Isophotal colors, NUVQTZ-50CCD and
FUVQTZ-50CCD, in the AB magnitude system. Photometry in the MAMA
images are discussed above. Photometry of objects falling partially
outside the MAMA image are flagged and should not be considered
reliable. When the measured flux is less than $2\sigma$, we give
lower limits to the color as discussed above.

{\bf $r_h$:} The half-light radius of the object in the 50CCD image,
given in milli-arcseconds. The half-light radius was determined by
{\sc SExtractor} to be the radius at which a circular aperture
contains half of the flux in the ``mag\_auto'' elliptical aperture.

{\bf s/g:} A star-galaxy classification parameter determined by a
neural network within {\sc SExtractor}, and based upon the morphology
of the object in the 50CCD images (see Bertin \& Arnouts \markcite{bertin96}1996 for a detailed
description of the neural network). Classifications near 1.0 are
more like a point source, while classifications near 0.0 are more
extended.

{\bf flags:} Flags are explained in the table notes, and include both
the flags returned by {\sc SExtractor}, and additional flags we
added while constructing the catalog.

\section{Statistics}

In this section we present several statistics of the data compiled
from the catalog.

\subsection{Source Counts}

\begin{figure}

\plotone{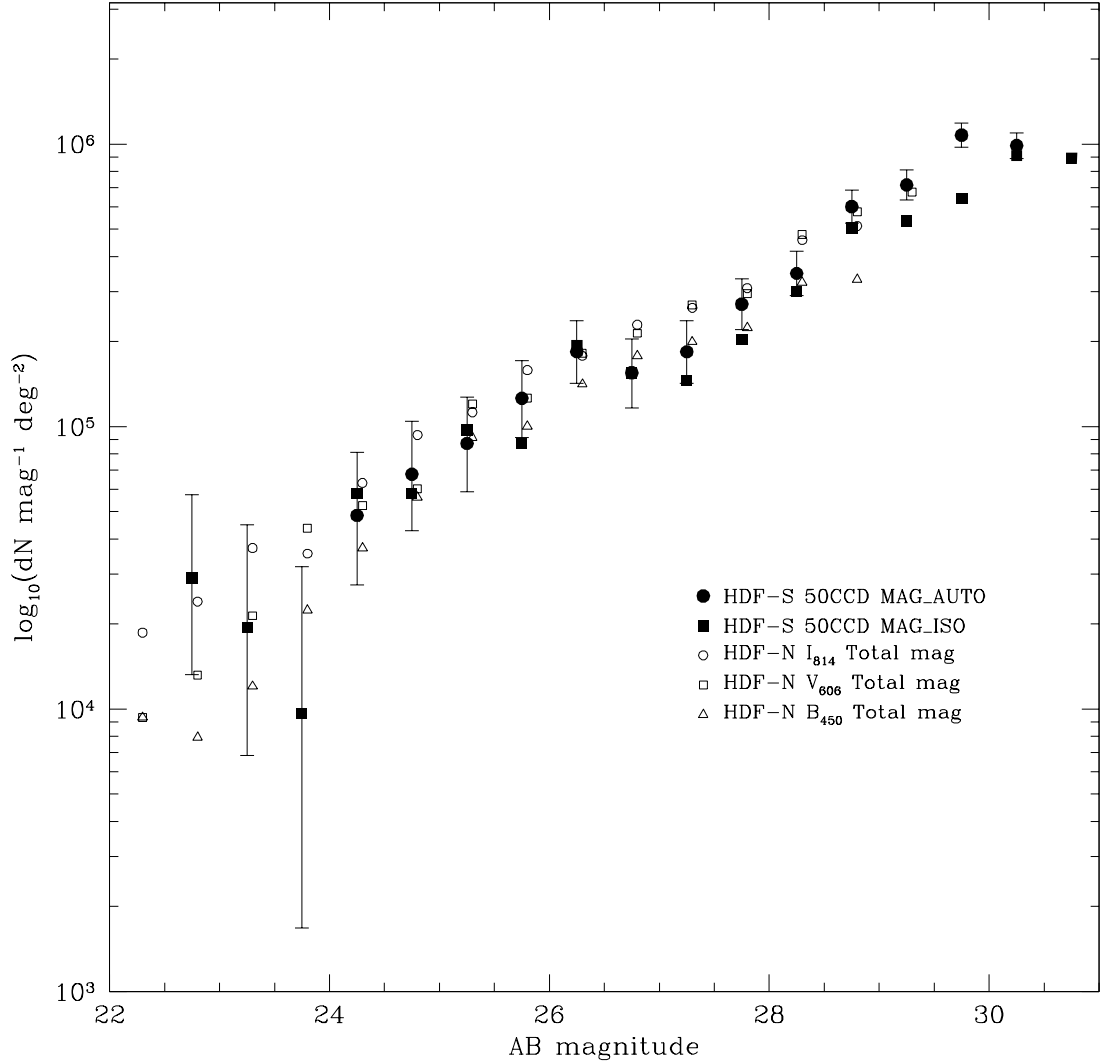}

\caption{The source counts in the 50CCD image scaled to objects
per square degree per magnitude as a function of AB magnitude. We
plot both the {\sc mag\_auto} and {\sc mag\_iso} counts, binned at
different magnitudes to show the points. We plot Poissonian errors
on the points. For comparison, we plot the WFPC2 HDF-N galaxy
counts, in $B_{450}$, $V_{606}$, and $I_{814}$, based upon the
total magnitude as given by Williams et al.\ \protect\markcite{williams96}(1996). The error
bars reflect $\sqrt{N}$ counting statistics and do not include
systematic errors in the photometry or galaxy clustering.}

\label{numcts}

\end{figure}

The source counts in the 50CCD image are given in Table~\ref{nctable},
and plotted as a function of AB magnitude in Figure~\ref{numcts},
where they are compared with the galaxy counts from the HDF-N WFPC2
observations, as compiled by Williams et al.\ \markcite{williams96}(1996). The counts are
compiled directly from the catalog, although all flagged regions
have been excluded, so that the counts do not include objects near
the edge of the image, or near the quasar. We plot only the Poissonian
errors, although there might be an additional component due to
large-scale structure. We plot all sources, including both galaxies
and stars, although we do not expect stars to contribute substantially
to the source counts. No corrections for detection completeness
have been made, and the counts continue to rise until fainter than
30~mag. The turnover fainter than this is due to incompleteness;
the counts do not turn over for astrophysical or cosmological
reasons. 

\subsection{Colors and Dropouts}

\begin{figure}

\plotone{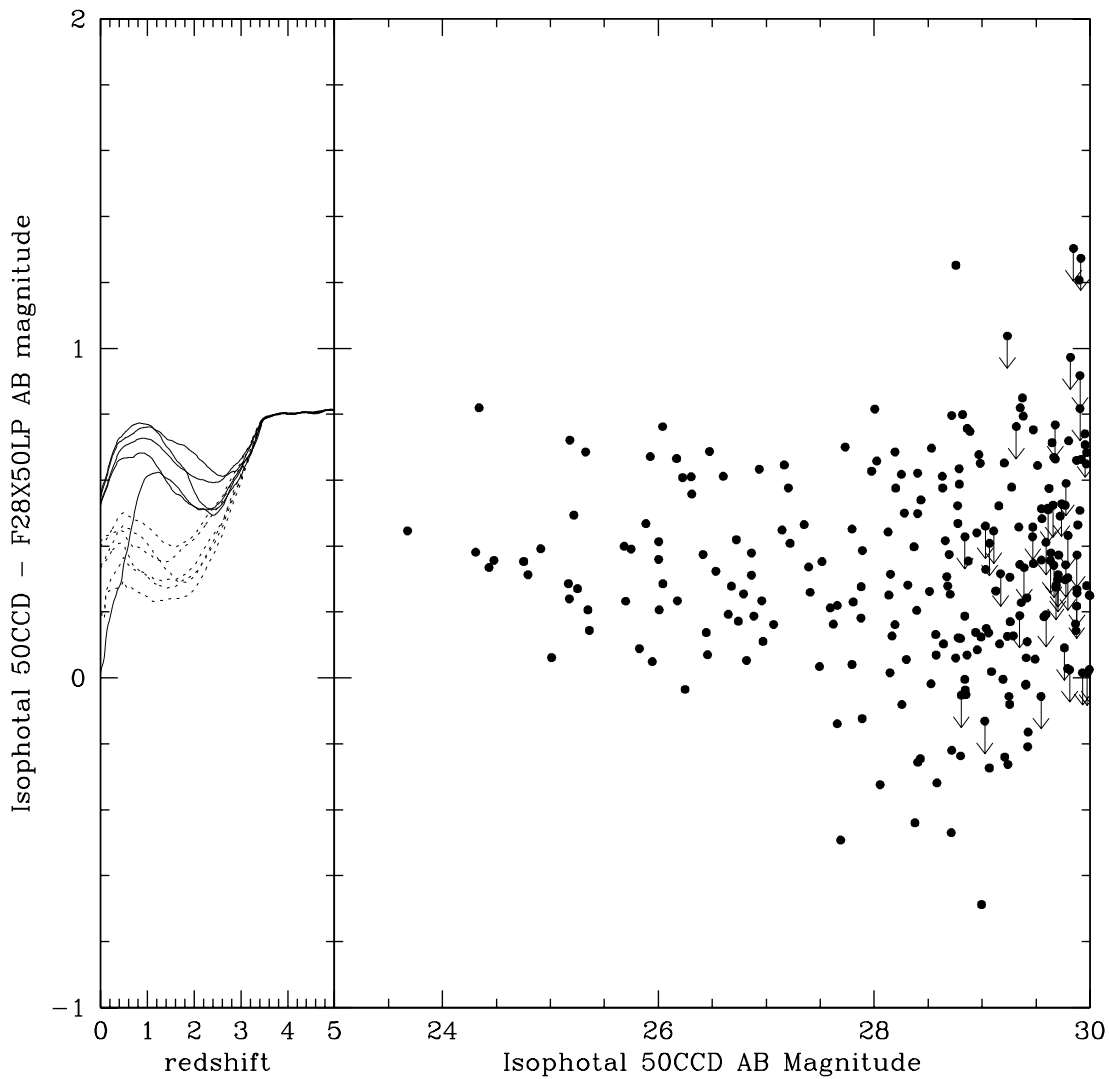}

\caption{50CCD-F28X50LP AB magnitudes plotted as a function of
50CCD magnitude. The magnitudes and colors are isophotal. On the
left we plot the K--corrected colors of the template galaxies in
the Kinney et al.\ \protect\markcite{kinney96}(1996) sample as a function of redshift. The
``normal'' galaxies from that sample are plotted as solid lines,
and the starburst galaxies are plotted as dotted lines. The templates
do not include data shortward of Ly$\alpha$, so the plots converge
when this limit is redshifted into the F28X50LP filter. In real
high-z galaxies, a similar effect would be caused by the Ly$\alpha$
forest.}

\label{lpcolor}

\end{figure}

The 50CCD-F28X50LP colors of objects in the STIS images are plotted
as points in Figure~\ref{lpcolor}. Flagged objects have been removed
from the sample. For comparison, we plot K--corrected (no-evolution)
colors of the template galaxies in the Kinney et al.\ \markcite{kinney96}(1996) sample
as a function of redshift on the left of the figure. The LP filter is
able to distinguish blue galaxies at $z<2.5$, but becomes
dominated by the noise for blue galaxies fainter than 28~mag, and
loses color resolution at $z>3$, where the Ly$\alpha$ forest
dominates the color in these bandpasses.

\begin{figure}

\plotone{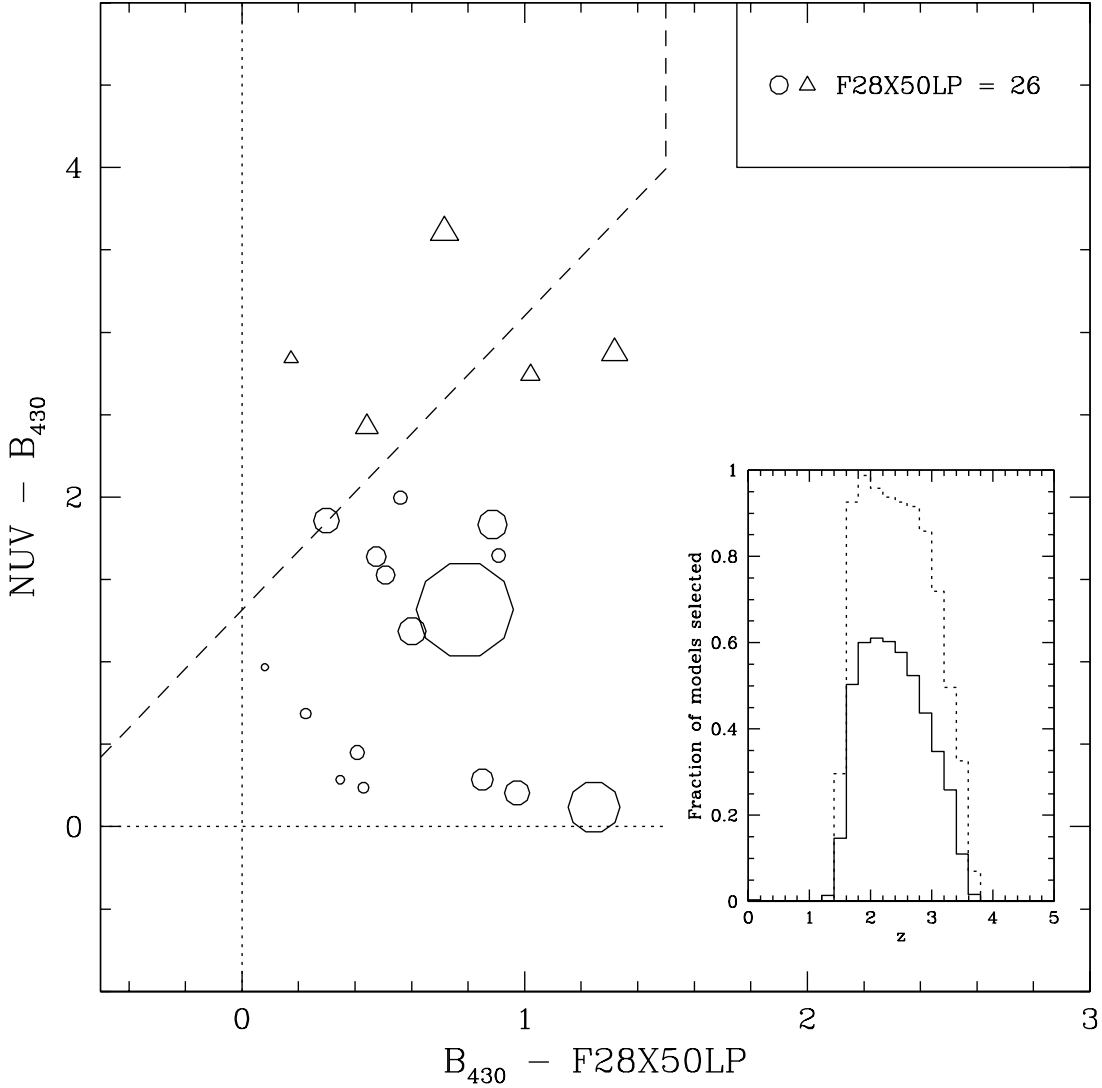}

\caption{A color-color plot of the STIS NUV - $B_{430}$ vs. $B_{430}$
- F28X50LP, where $B_{430}$ is a pseudo bandpass obtained by
subtracting a scaled F28X50LP flux from the 50CCD flux. The dashed
line shows the selection boundary for objects with $1.5<z<3.5$.
The size of the symbols indicates their magnitudes, and the symbol
size of an object with F28X50LP$~=~26$ is indicated in the inset
at the upper right. Circular symbols are detected at the 1 sigma
level in all bands, while triangles are undetected in the NUV,
providing lower limits to the color. In the inset figure at right,
we plot the selection efficiency of the NUV drop-out technique.
This shows the fraction of models from the Madau et al.\ \protect\markcite{madau96}(1996)
grid meeting the color selection criteria. The selection criteria
are NUV$-B_{430} > 1.75 (B_{430} - F28X50LP)+1.3$, AND $B_{430} -
F28X50LP < 1.5$. The solid line is the full set of models (including
old and highly reddened galaxies). The dotted line is just those
model galaxies with ages $<10^8$ years and foreground-screen
extinction less than $A_B = 2$.}

\label{nuvdrop}

\end{figure}

Because the F28X50LP bandpass is entirely contained within the
50CCD bandpass, it is possible, by subtracting an appropriately
scaled version of the measured F28X50LP flux from the 50CCD flux,
to construct a pseudo-$B_{430}$ measurement (see Figure~\ref{filttrans}).
This pseudo-$B_{430}$ is combined with the NUVQTZ and the F28X50LP
measurements in a color-color diagram in Figure~\ref{nuvdrop}. NUV
drop-outs, indicated on this figure by the dashed line, are those
objects with blue colors in the visible, but red colors in the UV,
indicative of galaxies at $z>\sim 1.5$. These galaxies show blue
colors characteristic of rapid star formation, while the red NUV
to optical color is due to the Lyman break and absorption by the
Ly$\alpha$ forest. The selection criteria were determined using
the models of Madau et al.\ \markcite{madau96}(1996). In an inset to Figure~\ref{nuvdrop},
we plot the efficiency of these criteria for selecting galaxies of
high redshift. The solid line is the fraction of all of the models
that meet these criteria, while the dotted line is the fraction of
those models with ages $<10^8$ years and foreground-screen extinction
less than $A_B = 2$. These criteria are very efficient at finding
young, star-forming galaxies at $1.5 < z < 3.5$. We have removed
point sources from this figure, including the bright object just
west of the QSO, which is extremely red and is likely to be an M
star.

\begin{figure}

\plotone{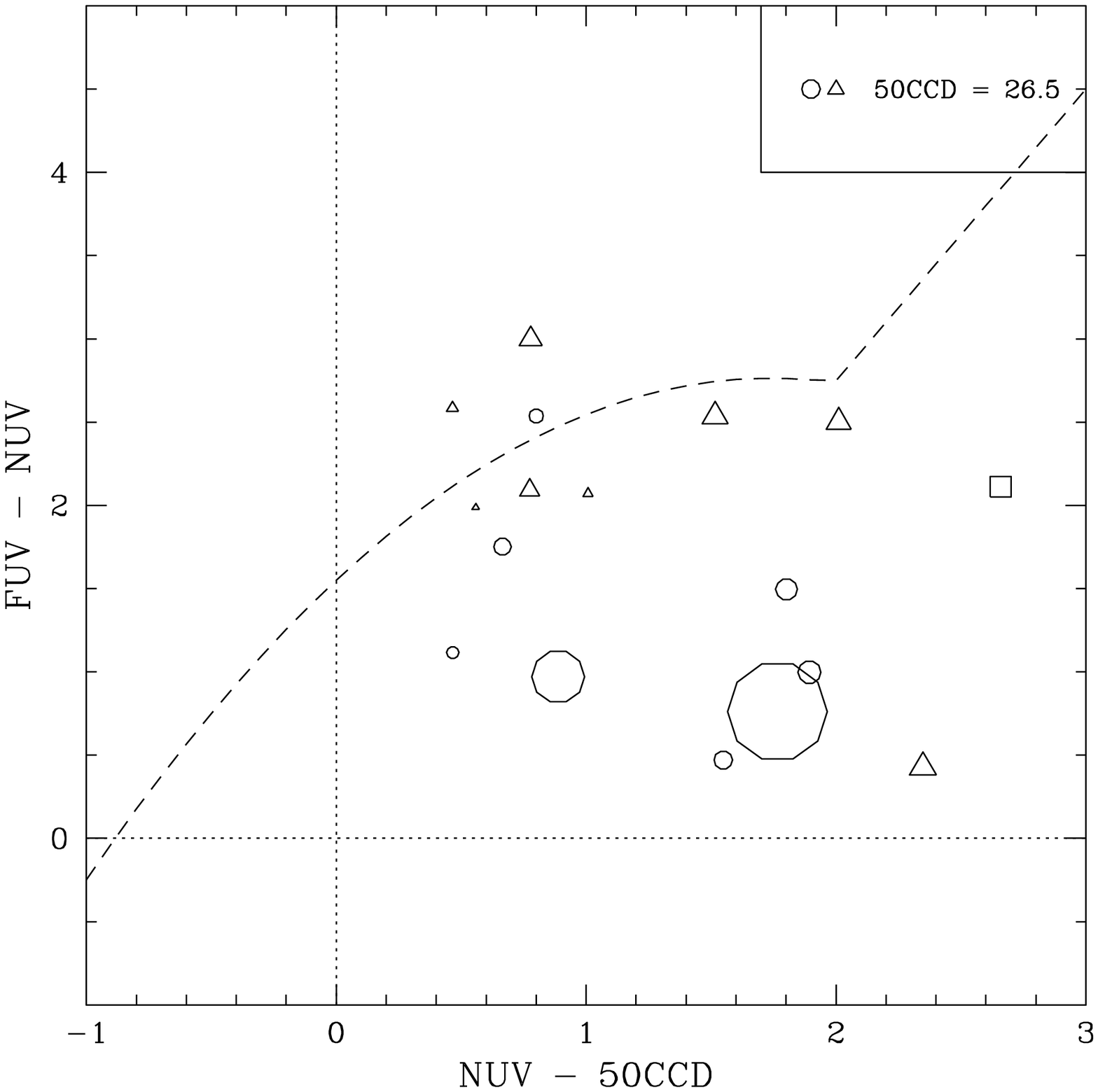}

\caption{A color-color plot of the STIS FUV - NUV vs. NUV - 50CCD.
The dashed line shows the selection boundary for objects with
$0.6<z<1.5$. The symbols are as in Figure~\protect\ref{nuvdrop};
in addition, the square symbols represent objects which are not
detected at the 1$\sigma$ level in either the NUV or the FUV.}

\label{fuvdrop}

\end{figure}

In Figure~\ref{fuvdrop} we give a FUV-NUV vs NUV-50CCD color-color
plot showing FUV dropouts, where the Lyman break is passing through
the FUV bandpass at $z>0.6$. Of the 17 galaxies in the MAMA field with
NUV magnitudes brighter than 28.4, only 3 have a clear signature of a
Lyman break at $0.6<z<1.5$. However, the upper limits are sufficiently
weak that roughly half the sample could be at $z>0.6$.

\section{Conclusions}

We have presented the STIS imaging observations that were done as
part of the Hubble Deep Field -- South campaign. The 50CCD image is
the deepest image ever made in the UV-optical wavelength region,
and achieves a point source resolution near the diffraction limit
of the HST. We have presented the catalog, and some statistics of
the data. These data will be useful for the study of the number and
sizes of faint galaxies, the UV-optical color evolution of galaxies,
the number of faint stars and white dwarfs in the galactic halo, and
the relation between absorption line systems seen in the QSO spectrum
and galaxies near to the line of sight. Follow-up observations of
the HDF-South fields by southern hemisphere ground-based telescopes,
by HST, and by other space missions will also greatly increase our
understanding of the processes of galaxy formation and evolution.

The images and catalog presented here are available on the World Wide
Web at: $<$http://www.stsci.edu/ftp/science/hdfsouth/hdfs.html$>$.

\bigskip

\acknowledgments

We would like to thank all of the people who contributed to making the
HDF-South campaign a success, including those who helped to identify a
target quasar in the southern CVZ, and those who helped in planning
and scheduling the observations. JPG, TMB, and HIT wish to acknowledge
funding by the Space Telescope Imaging Spectrograph Investigation
Definition Team through the National Optical Astronomical Observatories,
and by the Goddard Space Flight Center. CLM and CMC wish to acknowledge
support by NASA through Hubble Fellowship grants awarded by STScI.

\clearpage

\clearpage


}

\clearpage

\centerline{Table 2---Continued}

{\scriptsize
NOTE --- Flags: 
a: Object has near neighbors (50CCD). 
b: Object was orignally blended with another (50CCD). 
c: At least one pixel is saturated (50CCD). 
d: Object's aperture data is incomplete or corrupted (50CCD). 
e: Object is off the image, or within 30 pixels of the edge (50CCD). 
f: Object lies on top of a diffraction spike (50CCD). 
g: Photometry is from the second run of {\sc SExtractor} (50CCD). 
h: Object is off the image, or within 30 pixels of the edge (F28X50LP). 
i: Object lies on top of a diffraction spike (F28X50LP). 
j: Photometry is from the second run of {\sc SExtractor} (F28X50LP). 
k: Object is completely off the image (NUV). 
l: Object is partially off the image (NUV). 
m: Used global sky value (NUV). 
n: Object is completely off the image (FUV). 
o: Object is partially off the image (FUV). 
p: Used global sky value (FUV). 
q: Target QSO J2233-606.
}

\clearpage

{\scriptsize
\begin{deluxetable}{lrrrr}
\tablecaption{Source Counts}

\tablehead{
\colhead{AB mag} &
\colhead{N$_{auto}$} &
\colhead{log(N$_{auto}$)} &
\colhead{N$_{iso}$} &
\colhead{log(N$_{iso}$)}
}

\startdata
22.75&3&4.46&1&3.99\nl
23.25&2&4.29&2&4.29\nl
23.75&2&4.29&3&4.46\nl
24.25&5&4.68&1&3.99\nl
24.75&7&4.83&5&4.68\nl
25.25&9&4.94&12&5.07\nl
25.75&13&5.10&8&4.89\nl
26.25&19&5.26&15&5.16\nl
26.75&16&5.19&18&5.24\nl
27.25&19&5.26&17&5.22\nl
27.75&28&5.43&17&5.22\nl
28.25&36&5.54&23&5.35\nl
28.75&62&5.78&42&5.61\nl
29.25&74&5.86&55&5.73\nl
29.75&111&6.03&56&5.73\nl
30.25&102&5.99&92&5.95\nl
30.75&72&5.84&82&5.90\nl
31.25&27&5.42&92&5.95\nl
31.75&1&3.99&63&5.79\nl
32.25&1&3.99&8&4.89\nl

\tablecomments{The 50CCD source counts. Magnitudes are in the AB
magnitude system. The counts are given for the {\sc SExtractor}
{\sc mag\_auto}, which uses an elliptical aperture set at 2.5 times
the first moment of the semi-major and semi-minor axes. We also
give the counts as determined using isophotal magnitudes. Column 1 is
the center of the magnitude bin, columns 2 \& 4 contain the raw number
of objects per bin in the image, and columns 3 \& 5 contain the log of
the number per magnitude per square degree.}

\label{nctable}

\enddata
\end{deluxetable}
}
\end{document}